\def\useextern{}
\documentclass[conference,final]{IEEEtran}
\IEEEoverridecommandlockouts
\usepackage{ifluatex}
\ifluatex
\else
\usepackage[T1]{fontenc}
\usepackage[utf8]{inputenc} %support umlauts in the input
\fi

\usepackage{microtype}
\DisableLigatures{encoding = *, family = tt*}

\usepackage{amssymb}

\usepackage{graphicx}
\usepackage{comment}
\usepackage[hidelinks]{hyperref}
\usepackage[nameinlink,noabbrev,capitalize]{cleveref}
\usepackage{cite}

\ifluatex
\usepackage[newfloat,finalizecache]{minted}
\else
\usepackage[newfloat,frozencache]{minted}
\fi
\newminted{c}{}
\newminted{cpp}{}
\newminted{text}{}
\newmintinline{c}{style=bw}
\newmintinline{cpp}{style=bw}
\newmintinline{text}{style=bw}

\ifdefined\useextern

\usepackage{tikzexternal}
\else
\ifdefined\genextern

\usepackage{tikz}
\usetikzlibrary{external}
\else

\usepackage{tikz}

\fi

\usetikzlibrary{shapes}
\usetikzlibrary{fit}
\usetikzlibrary{calc}
\usetikzlibrary{bending}
\usetikzlibrary{arrows.meta} 
\usetikzlibrary{graphs}
\usetikzlibrary{decorations} 
\usetikzlibrary{decorations.text}
\usetikzlibrary{decorations.pathmorphing}
\usetikzlibrary{decorations.markings}
\usetikzlibrary{shadows.blur}

\usetikzlibrary{graphdrawing}
\usegdlibrary{layered}

\usepackage{pgfplots}
\pgfplotsset{compat=1.10}

\fi

\input{macros}

\begin{document}

\title{Autotuning Search Space for Loop Transformations}

\author{%
\IEEEauthorblockN{Michael Kruse, Hal Finkel}
\IEEEauthorblockA{%
\textit{Argonne Leadership Computing Facility}\\
\textit{Argonne National Laboratory}\\
Lemont, IL 60439, USA\\
\{mkruse,hfinkel\}@anl.gov}
\and
\IEEEauthorblockN{Xingfu Wu}
\IEEEauthorblockA{%
\textit{Mathematics \& Computer Science Division}\\
\textit{Argonne National Laboratory}\\
Lemont, IL 60439, USA\\
xingfu.wu@anl.gov}
}

\maketitle

\begin{abstract}
One of the challenges for optimizing compilers is to predict whether applying an optimization will improve its execution speed.
Programmers may override the compiler's profitability heuristic using optimization directives such as pragmas in the source code.
Machine learning in the form of autotuning can assist users in finding the best optimizations for each platform.

In this paper we propose a loop transformation search space that takes the form of a tree, in contrast to previous approaches that usually use vector spaces to represent loop optimization configurations.
We implemented a simple autotuner exploring the search space and applied it to a selected set of PolyBench kernels.
While the autotuner is capable of representing every possible sequence of loop transformations and their relations, the results motivate the use of better search strategies such as Monte Carlo tree search to find sophisticated loop transformations such as multilevel tiling.
\end{abstract}

\begin{IEEEkeywords}
autotuning, Clang, Polly, loop transformation, performance optimization
\end{IEEEkeywords}

% !TeX encoding = UTF-8
% !TeX spellcheck = en_US
% !TeX root = paper.tex

%%%%%%%%%%%%%%%%%%%%%%%%%%%%%%%%%%%%%%%%%%%%%%%%%%%%%%%%%%%%%%%%%%%%%%%%%%%%%%%%
\section{Motivation}

Most compute-intensive programs, in particular scientific applications in high-performance computing, spend a significant amount of execution time in loops, thus making them the prime target for performance optimizations.
Common strategies to speed up such programs include replacing runtime-dominating kernels with a preoptimized function calls to a library often provided by the target platform vendor, analyzing and applying best practices or hints by tools such as Intel V-Tune, or just trying out different loop optimization and choosing the one that turns out to be the best performance.

Unfortunately, preoptimized libraries do not provide functions for every algorithm and often require programs to match the data structures expected by the library.
On the other hand, hand-optimizing the program requires a lot of engineering time and needs to be redone for every target hardware.
That is, every new generation of computer hardware requires maintenance work, which also applies to preoptimized libraries themselves.

A third approach is to automate the search for the best-performing program, called \emph{autotuning}.
If the search space is small enough, one can try every variation of a program and keep the fastest to use for production runs.
 Usually, however, the search space is too large to search exhaustively. Instead,  machine learning is used to find the best candidates while evaluating only a subset of program variants. 

An often-used search space is the set of compiler command line flags that results in the fastest program~\cite{fursin11-milepost,ansel14-opentuner}. This can be different from using the default high optimization setting (``\textinline{-O3}'') or just enabling every supported optimization.
The advantage of this approach is that it is easy to implement and generally applicable.
This approach is not very powerful, however, since  it cannot profit from the program structure or domain knowledge and applies the same flags to all code in a translation unit.
For instance, passing the internal option \textinline{-mllvm -unroll-count=4} to Clang will unroll every loop by a factor of 4.

Understandably, the optimal unroll factor differs between loops.
Loop transformations such as unrolling can also be controlled by using directives in the source.
For instance, \cinline{#pragma unroll(4)} before the loop instructs the compiler to override its own heuristics to choose an unroll factor and use 4 instead.
Hence it can be configured separately for each loop.
Unfortunately, today's general-purpose compilers, such as gcc, clang, xlc, icc, msvc, and nvc, lack good support for such loop transformation directives apart from unrolling and vectorization.
For the loop transformations that are supported, each compiler uses a different syntax and subtly different semantics.
Furthermore, at most one transformation can be reliably applied to a loop.
% Stacking transformations, such as first interchange two loops and then vectorize the inner loop, is generally not supported.

We have been working on improved loop transformations for Clang/LLVM~\cite{llvmhpc18-pragmas}. 
In addition to the loop unrolling, unroll-and-jam, vectorization, and loop distribution pragmas already supported by Clang, we added tiling, loop interchange, reversal, array packing, and thread-parallelization directives.
Moreover, and maybe more important, these transformations can be composed. For instance, first a loop can be  tiled, and then the outer loop can be parallelized and the inner loop vectorized.

The set of applied loop transformations has structure itself that can be useful to incrementally build more sophisticated loop transformations.
An example is \cref{lst:gemmcore}, which optimizes a matrix-matrix multiplication using loop transformation pragmas.
It  follows a BLIS-like optimization inspired by~\cite{low16-blis}, but without a microkernel.

\begin{listing}
\begin{minted}[fontsize=\footnotesize]{c}
#pragma clang loop(i1) pack array(B) allocate(malloc)
#pragma clang loop(j2) pack array(A) allocate(malloc)
#pragma clang loop(i1,j1,k1,i2,j2) interchange \
    permutation(j1,k1,i1,j2,i2)
#pragma clang loop(i,j,k) tile sizes(448,2048,256) \
    floor_ids(i1,j1,k1) tile_ids(i2,j2,k2)

#pragma clang loop id(i)
for (int i = 0; i < _PB_NI; i++)
  #pragma clang loop id(j)
  for (int j = 0; j < _PB_NJ; j++) 
    #pragma clang loop id(k)
    for (int k = 0; k < _PB_NK; k++)
      C[i][j] += alpha * A[i][k] * B[k][j];
\end{minted}
\caption{Loop transformations applied to the deepest loop nest of a naive gemm implementation.}\label{lst:gemmcore}
\end{listing}

Generally, a loop transformation is applied to the loop on the next line, respectively the result of the loop transformation on the next line, such that textually the transformations appear in reverse order.
Here, however,  we assign unique identifiers to literal loops (\cinline{loop(i1)}) and loops that are generated by transformations (\cinline{tile_ids(i2,j2,k2)}), such that we can refer to multiple loops at the same time and disambiguate the transformation order.

The directives in \cref{lst:gemmcore} first apply tiling on a perfect loop nest consisting of the loops i,j,k, which results in a loop nest of six loops named i1,j1,k1,i2,j2,k2.
The interchange directive next changes the nesting order to j1,k1,i1,j2,i2,k2 (the loop k2 remains the innermost loop and thus does not participate in the permutation).
Then, the working sets within the loops j2 and i1 of the arrays A and B are copied to fit into the L2 and  the L1 caches, respectively. 

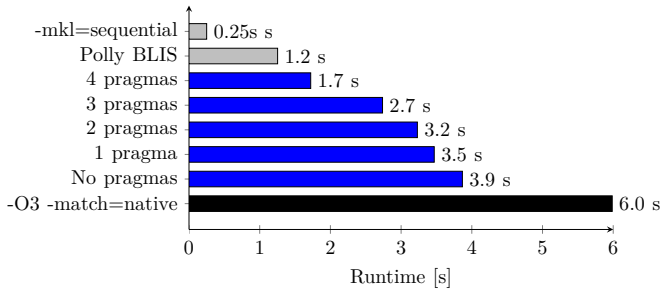
\begin{figure}
\begin{resizepar}
\begin{tikzpicture}
\begin{axis}[
xbar,width=90mm,bar width=1.8ex,bar shift=0pt,height=55mm,
ytick=data,
symbolic y coords={
base,
nothing,
tile,
tile_interchange,
tile_interchange_packA,
tile_interchange_packAB,
polly,
mkl},
yticklabels={{-O3 -match=native},{No pragmas},{1 pragma},{2 pragmas},{3 pragmas},{4 pragmas},{Polly BLIS},{-mkl=sequential}},
axis y line=left,
enlarge y limits=0.15,           % Y-Axis
xmin=0,xmax=6,xlabel={Runtime [s]},axis x line=bottom,    % X-Axis
nodes near coords={\textcolor{black}{\pgfplotspointmeta}},point meta=explicit symbolic,nodes near coords align={horizontal},nodes near coords style={right} %
]

\addplot[draw=none,fill=none] coordinates {
% MKL sequential 0.249785
% Polly: 1.253210
% tile+interchange+pack(A,B): 1.721389
% tile+interchange+pack(A): 2.737203
% tile+interchange: 3.230106
% tile: 3.468643
% Polly nothing: 3.866791
% -O3: 5.985794
(0,base) 
(0,nothing) 
(0,tile) 
(0,tile_interchange) 
(0,tile_interchange_packA) 
(0,tile_interchange_packAB)
(0,polly)
(0,mkl) 
};

\addplot[draw=black,fill=gray!50] coordinates {
(1.253210,polly) [1.2 s]
(0.249785,mkl) [0.25s s]
};

\addplot[draw=black,fill=blue] coordinates {
(3.866791,nothing) [3.9 s]
(3.468643,tile) [3.5 s]
(3.230106,tile_interchange) [3.2 s]
(2.737203,tile_interchange_packA) [2.7 s]
(1.721389,tile_interchange_packAB) [1.7 s]
};

\addplot[draw=black,fill=black] coordinates {
(5.985794,base) [6.0 s]
};

\end{axis}
\end{tikzpicture}
\end{resizepar}
\caption{Comparison of matrix-matrix multiplication performance  with that of a baseline configuration.}\label{fig:gemm_successive}
\end{figure}

\Cref{fig:gemm_successive} shows how adding more pragmas successively improves the kernel's performance.
\dquote{1 pragma} applies only the \cinline{tile} pragma from \cref{lst:gemmcore}; \dquote{2 pragmas} adds the \cinline{interchange} directive; and so on.
All except the baseline also have Polly enabled, which in our implementation applies the loop transformations.
This is why \dquote{No pragmas} differs in performance: even though no explicit transformation is applied, Polly applies its polyhedral modeling and lowers it back to LLVM-IR.
Part of the process is applying loop-versioning and additional no-aliasing metadata to the IR, which is helpful for LLVM's inner loop vectorizer.
That is, vectorization is implicitly applied as well as a fifth transformation.

% TODO: parallelize

For comparison, \cref{fig:gemm_successive} also shows the runtime when using Polly's built-in BLIS optimization~\cite{gareev18-gemm}, which is also modeled after~\cite{low16-blis} but additionally applies microkernel optimizations that we currently cannot replicate using pragma directives.
For reference, we also include the runtime when using Intel MKL's \cinline{cblas_dgemm} implementation, which can be considered the fastest matrix-matrix multiplication implementation on that machine.

This motivates a treelike search space for loop transformations: successively apply transformations to existing loops or loops generated by previous transformations to gradually improve performance.
We explore this treelike search in this article.

This motivation might make a tree-like search space appear fairly obvious, yet autotuning using Monte-Carlo tree search has been implemented only recently~\cite{hajali20-protuner} and published after we started investigating this approach.
We suspect the dominance of machine learning tools --- including those used for autotuning --- on vector spaces is one of the reasons.

%%%%%%%%%%%%%%%%%%%%%%%%%%%%%%%%%%%%%%%%%%%%%%%%%%%%%%%%%%%%%%%%%%%%%%%%%%%%%%%%
\section{Our Contributions}

We summarize the contribution of this paper with the following items.

\begin{itemize}
\item The idea of representing the loop transformation search space as a tree by stacking up more and more loop transformations.
\item A simple implementation of an autotuner using this search space: \emph{mctree}.
\item An evaluation of the autotuner on a selection of loop nests.
\end{itemize}
To avoid confusion, we explicitly declare the following as \emph{non-goals} of this paper:
\begin{itemize}
\item Find new implementation techniques for the selected algorithms (gemm, syr2k, covariance).
    The loop nests were selected because their performance characteristics are well known and existing implementations can be compared against as a baseline.
    However, autotuning might help reduce the engineering effort needed to get close to the optimum/baseline, especially for algorithms that do not get as much attention.
\item Design and implement the loop transformation pragmas. 
    These have been the topic of our previous publications~\cite{llvmhpc18-pragmas} and~\cite{iwomp18-pragmas}.
\item Implement an autotuning framework. 
    There are dedicated projects such as Collective Knowledge~\cite{fursin18-ck}, OpenTuner~\cite{ansel14-opentuner}, ATF~\cite{rasch17-atf}, and ytopt~\cite{ytopt}.
\item Develop a machine learning strategy for autotuning.
    We refer to~\cite{balaprakash18-surf,ashouri18-survey} for an overview of search space exploration techniques.
\end{itemize}

%%%%%%%%%%%%%%%%%%%%%%%%%%%%%%%%%%%%%%%%%%%%%%%%%%%%%%%%%%%%%%%%%%%%%%%%%%%%%%%%
\section{Loop Transformation Search Space}\label{sct:searchspace}

% Ideally, every possible loop transformation combination is representable in the search space and reflects the problem structure such that the best performing configurations are straightforward to explore and simpler solutions are easier to find than complicated ones.
% TODO: How does our search space fulfill this?

The baseline without any transformations serves as the root node from which all other configurations are derived by adding transformations.
As a result, the search space has the structure of a tree.

For deriving a new configuration, compute the loop structure after applying the previous configurations.
For instance, after applying the first pragma from the previous matrix-matrix multiplication example, the loop nest structure has six instead of three loops.
\begin{minted}[fontsize=\footnotesize]{c}
for (int i1 = 0; i1 < _PB_NI; i1+=448)
  for (int j1 = 0; j1 < _PB_NJ; j1+=2048) 
    for (int k1 = 0; k1 < _PB_NK; k1+=256)
      for (int i2 = i1; i2 < i1+448; i2++)
        for (int j2 = j1; j2 < j1+2048; j2++) 
          for (int k2 = k1; k2 < k1+256; k2++)
            C[i2][j2] += alpha * A[i2][k2] * B[k2][j2];
\end{minted}
More transformations can be applied after the tiling as if the source code had six loops in the first place.
The case that the loop trip counts are not multiples of the tile sizes is handled transparently by the compiler.
It may, for example, add remainder loops that are \squote{hidden} from the loop structure in order not to  interfere with the perfect nesting that could inhibit the application of further transformations; or it may add conditional execution/masking into the loop nest body.
The rationale is that the execution of remainder loops is usually not the performance-critical part compared with the complete tiles. 
%An idea for future work is to make the handling of remainder loops as a transformation option.

The children of a search tree node include all transformations that structurally can be applied to the loop nest. 
For instance, the perfectly nested loop nest of the matrix-matrix multiplication with three loops can support a tiling transformation for up to three dimensions, such as a two-dimensional tiling of the \cinline{i} and \cinline{j} loops.
Additionally, the tiling requires a size for each dimension, for which we chose the Cartesian product of a predefined set of tile sizes.

\begin{figure}
\includegraphics[width=\linewidth]{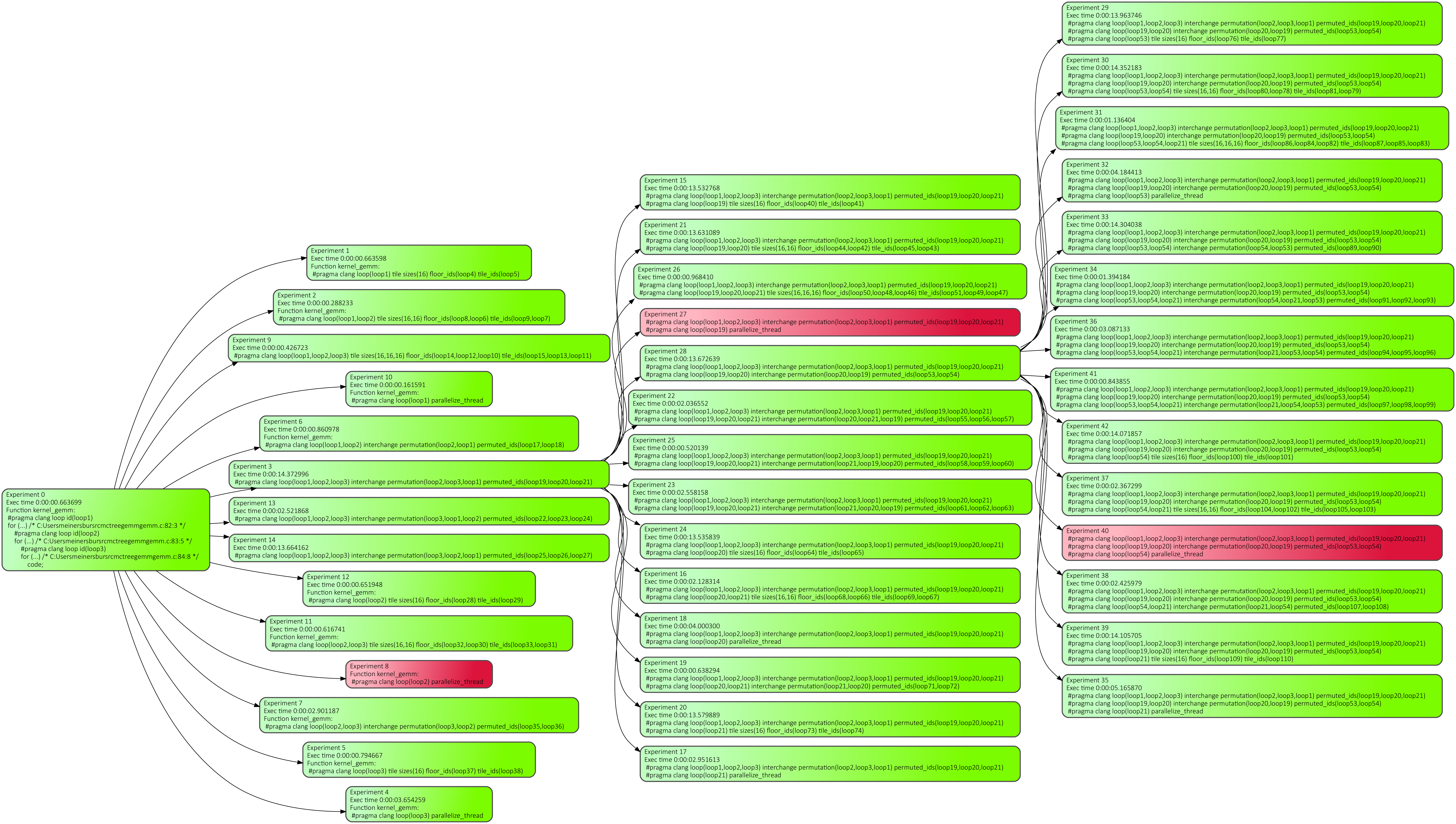}
\caption{Output of autotuning with a loop nest with 3 loops limited to a single tile size after 3 node expansions. Configurations where either the compilation or execution failed are red. Successful executions are marked green.}\label{fig:mctree_dot}
\end{figure}

Depending on the transformation, applying it multiple times can make sense, such as tiling for multiple caches in the memory hierarchy.
It follows that the search space is conceptually infinite.
An example of a partially expanded search space using interchange and tiling with a single tile size is shown in \cref{fig:mctree_dot}.

Some transformations, such as thread-parallelization, do not make sense to be applied multiple times: once a loop is parallelized, there is no meaning in parallelizing it again.
It can make sense to prune the search space before trying to evaluate a configuration or derive more configurations from it.

In other cases, only the compiler applying the transformations knows whether a transformation can be applied.
For instance, to determine whether a transformation is semantically legal, the compiler has to apply a dependency analysis.
While one can implement this in the search space creator as well, the compiler is much better suited for this analysis.
In \cref{fig:mctree_dot}, configurations that are rejected by the compiler are highlighted in red.

One can reach the same configuration through multiple paths. 
The most obvious case is with two loops: a transformation that does not affect other loops is first applied on one loop and then the other, respectively, in reverse order.
Less obvious cases are where a transformation can be understood as a shortcut of two transformations. For example, a partial unrolling by a factor has essentially the same effect as first tiling the loop by that factor and then fully unrolling  the inner loop, which now has a constant trip count. 
Smaller differences may occur due to different handling of remainder iterations.
The point is that the search tree is actually a directed acyclic graph, unless we allow for the same configuration to appear multiple times in the tree.

Transformations that have parameters contribute significantly to the number of children, such as the different choices of tile sizes.
For this reason, deriving configurations in multiple steps might be beneficial: first applying a transformation in its generic form and then deriving variations of its parameters.
The parameters do not necessarily need to be an exhaustive list of all possibilities but can also use strategies such as Bayesian prediction or simulated annealing that exploit the numeric structure not found in the tree structure.

%%%%%%%%%%%%%%%%%%%%%%%%%%%%%%%%%%%%%%%%%%%%%%%%%%%%%%%%%%%%%%%%%%%%%%%%%%%%%%%%
\section{Implementation}

We implemented a simple demonstration\footnote{\url{github.com/Meinersbur/mctree}} in Python and slightly extended our implementation\footnote{\url{github.com/Meinersbur/llvm-project/tree/pragma-clang-loop}} of loop transformation directives from~\cite{llvmhpc18-pragmas}.

The implementation can be thought of as consisting of three components:
the compiler for finding loop nests and applying transformations,
the search space generator for deriving new configurations,
and the autotuning driver for choosing configurations and executing commands.
We discuss these components in detail in the following sections.

\subsection{Loop Nest Analysis and Transformation: Clang/Polly}

\begin{figure}
\begin{resizepar}
\begin{tikzpicture}
		\tikzset{tight/.style={inner sep=0pt,outer sep=0pt,minimum size=0pt}}
		\tikzset{node/.style={draw,fill=white,line width=1.2pt,rounded corners,drop shadow}}
		\tikzset{supernode/.style={subgraph text none,draw,rounded corners}}
		\tikzset{edge/.style={->}}
		\graph[layered layout,edges={edge,rounded corners},level sep=5mm,sibling sep=10mm]{
			c[as={\\\texttt{\#pragma clang loop}\\\texttt{for (int i=...)}\\\hspace*{4mm}\dots},align=flush left,grow=right,draw,shape=file,fill=white,label={source.c}];
			ir[as={IR},shape=file,draw,fill=white,nudge=(up:10mm)];
			asm[as={Assembly},shape=file,draw,fill=white];
			json[as={\texttt{loopnests.json}},shape=file,draw,fill=white];
			
			clang [subgraph text none,label={[font=\Large\sffamily]above:Clang}] // [sibling sep=2mm,grow=down,layered layout] {
				lexer [as={Lexer},node];
				parser [as={Parser},node,grow=down];
				preprocessor [as={Preprocessor},node];
				sema [as={Semantic Analyzer},node];
				codegen [as={IR Generation},node];
				lexer->preprocessor->parser->sema->codegen;
			};
			llvm [subgraph text none,label={[font=\Large\sffamily]above:LLVM}] // [sibling sep=2mm,grow=down,layered layout] {
				canonicalization [as={Canonicalization passes},node];
				loopopts [as={Loop optimization passes},node];
				polly [as={Polly},node];
				vectorization [as={LoopVectorize},node];
				latepasses [as={Late Mid-End passes},node];
				backend [as={Target backend},node];
				canonicalization->loopopts;
				vectorization->latepasses->backend;
				loopopts->polly->vectorization;
			};
			c->[in=180]lexer;
			codegen->[out=0,in=-90]ir;
			ir->[out=90,in=180]canonicalization;
			backend->[out=0,in=-90]asm;
			polly->[edge node={node[above,sloped,font=\ttfamily\tiny]{-polly-output-loopnest}}]json;
		};
		\begin{pgfonlayer}{background}
		\node[tight,fit={(clang)},supernode]{};
		\node[tight,fit={(llvm)},supernode]{};
		\end{pgfonlayer}
		\path (preprocessor) edge[edge,densely dotted,bend left=50,draw=blue!50!black] node[midway,right,font=\ttfamily,blue!50!black] {\#pragma} (sema);
		\path (preprocessor) edge[edge,densely dotted,bend right=80,draw=blue!50!black] node[pos=0.7,left,font=\ttfamily,blue!50!black] {\#pragma} (codegen);
		\path (codegen) edge[edge,dashed,bend right=80,draw=blue!50!black,postaction={decorate,decoration={text along path,text={|\color{blue!50!black}|Loop metadata},raise=-1.7ex,pre=moveto,pre length=13mm}}] (ir);
		%\path (ir) edge[edge,dashed,draw=blue!50!black] (loopopts);
		\path (ir) edge[edge,dashed,draw=blue!50!black] (polly);
		%\path (ir) edge[edge,dashed,draw=blue!50!black,bend right=10] (vectorization);
		%\path (polly) edge[edge,dashed,draw=blue!50!black,bend right=10] (vectorization);
		%
\end{tikzpicture}
\end{resizepar}
\caption{Clang/LLVM compiler architecture}\label{fig:clangpolly}
\end{figure}
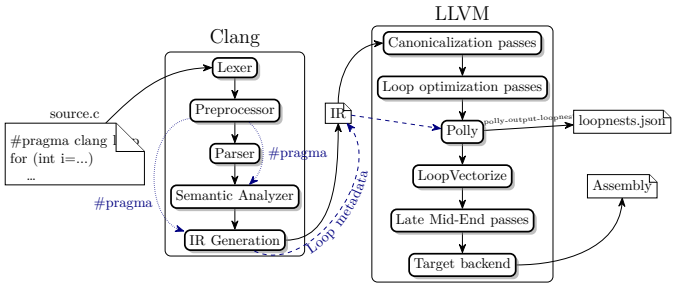

We already discussed our implementation of loop transformation directives in our implementation~\cite{llvmhpc18-pragmas}. 
To summarize, Clang/LLVM~\cite{lattner02-llvm,clang} follows the typical compiler structure with a front-end, mid-end, and back-end, illustrated in \cref{fig:clangpolly}.
Since \cinline{#pragma} are preprocessor directives, these are first processed by Clang's built-in preprocessor.
The annotation is forwarded to Clang's IR generation, which converts the pragmas into metadata that is attached to the loop that the pragma is associated with.

The IR including the metadata is passed to LLVM, the compiler's mid-end where most optimizations take place.
One of the optimization passes in LLVM is Polly~\cite{grosser12-polly}, LLVM's polyhedral loop nest optimizer.
Polly searches for loop nests compatible with the polyhedral model, which it can optimize by optimizing a proxy integer linear program.
However, the polyhedral representation also makes it straightforward to explicitly apply specific loop transformations.
Hence, our modification picks up the attached loop transformation metadata and applies it to the polyhedral representation, which then is converted back to LLVM-IR.

To be able to determine the loop nest structure that transformations can apply to, we added a new option \textinline{-polly-output-loopnest} that takes a filename to write to as an argument.
With this option, Polly emits a list of loop nest structures of all polyhedral representations found in a JSON format. 
 JSON also includes the source code locations of the loops when compiled with debug information.
Otherwise, debug locations are not emitted into the intermediate representation.

Plugging directly into Polly ensures that the search space is built from exactly the set of loop nests that can be transformed.
With the alternative of identifying loops at the AST level, the file might interpret every for-loop as transformable and miss cases such as while-loops and combined loop nests after inlining.

\subsection{Search Space Generation: mctree}\label{sct:mctree}

We implemented three different transformations in the search tree generator: tiling, loop interchange (also called loop permutation, which clarifies the handling of more than two loops), and thread-parallelization.
\begin{itemize}
\item Loop tiling: In addition to the number of loops to tile in a perfect loop nest, a new configuration is derived for each tile size from a preconfigured set of tiles sizes. 
For instance, with a perfect loop nest, the two loops \texttt{i} and \texttt{j}, and the tile sizes 2 and 4, the following six configurations are derived:
\begin{minted}[fontsize=\footnotesize]{c}
#pragma clang loop(i) tile sizes(2)
#pragma clang loop(i) tile sizes(4)
#pragma clang loop(i,j) tile sizes(2,2)
#pragma clang loop(i,j) tile sizes(2,4)
#pragma clang loop(i,j) tile sizes(4,2)
#pragma clang loop(i,j) tile sizes(4,4)
\end{minted}
The configurations using \texttt{j} as the outermost loop is generated as well, by interpreting \texttt{j} the outermost loop of the perfect loop nest.

\item Loop interchange: For each perfect loop nest,  a new configuration is derived for each permutation of its loops. 
For instance, for a loop nest of three loops, the following transformations are applied:
\begin{minted}[fontsize=\footnotesize]{c}
#pragma clang loop(i,j) interchange permutation(j,i)
#pragma clang loop(i,j,k) interchange permutation(j,k,i)
#pragma clang loop(i,j,k) interchange permutation(k,i,j)
#pragma clang loop(i,j,k) interchange permutation(k,j,i)
\end{minted}
The configurations where \texttt{i} remains the outermost loop are generated when \texttt{j} is interpreted as the outermost loop of a perfect loop nest.

\item Thread-parallelization: For each loop,  a new configuration is created by adding \cinline{#pragma clang loop parallelize_thread}.
The implementation will generate code equivalent to \cinline{#pragma omp parallel for schedule(static)} for such a loop, in other words, require an OpenMP runtime.
The reason for not using the OpenMP directive directly is that it is not composable with other loop transformations. 
In Clang, OpenMP lowering is implemented in the front-end, whereas loop transformations are implemented as mid-end optimization passes.
\end{itemize}

This limited set should still suffice to demonstrate autotuning on such a search space.
An interesting additional transformation would have been array packing, which would enable the autotuner to find the motivational gemm optimization of \cref{fig:gemm_successive}. 
However, this would additionally require the information that arrays are available in the input program. Unfortunately, in the intermediate representation, variable names are not  easy to recover after being lowered to virtual registers.

Internally, the loop nest is represented by an object tree; each object represents a loop and stores its unique name. 
After applying a transformation, the transformed loop objects are replaced with new ones representing the structure of the nest after the transformation.
That is, tiling $n$ loops removes those objects and reinserts twice as many in their place. 
Interchange reinserts the same numbers of loops.
Parallelization does not insert any new loop object, and an already parallelized loop is not considered to be any more transformable.
Loops not affected by a transformation keep their identifiers.

We did not implement any additional search pruning; instead we rely on Polly to reject any malformed transformation sequence.
Also, detection of equivalent transformations through different paths has not been implemented.

\emph{Monte Carlo tree search} was intended to be used as the search space exploration technique from the beginning, and hence the origin of the name \emph{mctree}.

\subsection{Autotuning Driver: mctree autotune}

The autotuner is integrated with the search space generator as a command line subcommand.
It takes a compiler command line as an argument from which it can extract which Clang executable to use, the input source files, and the output executable.
The command line must include the linking step, that is, must not contain the \textinline{-c} switch.

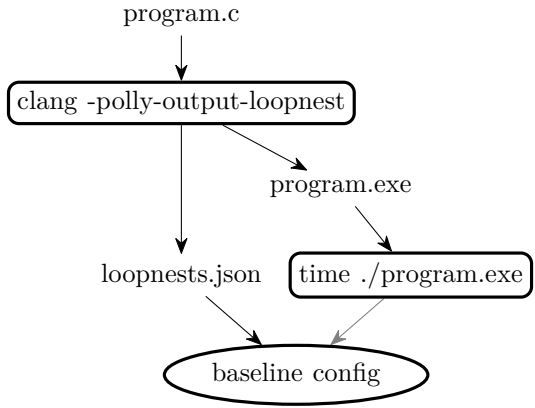
\begin{figure}
\begin{tikzpicture}[font=\sffamily]
\tikzset{node/.style={draw,line width=1.2pt,rounded corners}}
\tikzset{config/.style={draw,ellipse,line width=1.2pt,rounded corners}}

\graph[layered layout,grow=down,layer sep=6mm] {
input[as={\texttt{program.c}}];

base_compile[as={\texttt{clang -polly-output-loopnest}},node];
base_json[as={loopnests.json}];
base_exe[as={program.exe}];
base_run[as={time ./program.exe},node];
base[as={baseline config},config];

input -> base_compile;
base_compile -> base_json;
base_compile -> base_exe;
base_exe -> base_run;

base_json->base;
base_run->[gray]base;
};
\end{tikzpicture}\centering
\caption{Baseline configuration evaluation}\label{fig:baseline}
\end{figure}

As the first step it explores the baseline configuration (see \cref{fig:baseline}) by running the command line with the \textinline{-mllvm -polly-output-format} option added and enabling debug information.
The two outputs are the information of the loops structure in the form of a JSON file and an executable. 
The executable is executed to determine the runtime of the baseline configuration, since it might be the fastest configuration.
This is experiment number 0.
The JSON file is used to construct the loop nest structure of the root experiment from which all other configurations are derived, which is placed in a priority queue.

The fastest configuration whose children have not been explored yet is taken from the priority queue.
The rationale is that the configuration that was the fastest so far is also the most promising to derive new configurations from.
That is, we implemented only an extreme form of Monte Carlo tree search with exploitation only, but no exploration.
An alternative description could be hill climbing with backtracking.
Initially, the only configuration is the baseline configuration.

New configurations are derived from the selected configuration as previously described.
The execution time for each of the configurations is determined and then also inserted into the priority queue.
This process is repeated until no more elements are in the priority queue. 
The order of evaluation is also determined by the priority queue with its heap implementation, so the experiment number between sibling configurations is not meaningful.
Since the search space is potentially infinite, this situation may never happen, and the process has to be interrupted manually.

\begin{figure}
\begin{tikzpicture}[font=\sffamily]
\tikzset{node/.style={draw,line width=1.2pt,rounded corners}}
\tikzset{config/.style={draw,ellipse,line width=1.2pt}}

\graph[layered layout,grow=down,layer sep=4mm] {
input[as={\texttt{program.c}}];
config[as={derived config},config];

insertpragmas[as={insert directives},node];

{input,config}->insertpragmas;

withpragmas[as={\texttt{rewritten/program.c}}];
insertpragmas->withpragmas;

compile[align=flush left,as={\texttt{clang -fopenmp \textbackslash}\\\texttt{-Werror=pass-failed}},node];
exe[as={rewritten/program.exe}];
run[as={time rewritten/program.exe},node];
withpragmas->compile->exe->run;
%run->[no span edge]config;
};
\path (run) edge[->,bend right=50,gray,postaction={decorate,decoration={text along path,text={augment},raise=-1.7ex,pre=moveto,pre length=20mm}}] (config);
%\path (run) edge[->,bend right=30,gray,edge node={node[sloped,above,font=\footnotesize]{augment}}] (config);
\end{tikzpicture}\centering
\caption{Derived configuration evaluation}\label{fig:derived}
\end{figure}
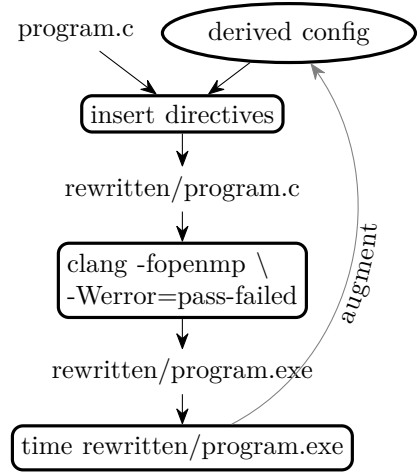

The execution time of derived configurations is determined as illustrated in \cref{fig:derived}. 
Using the source code location from the JSON file, each loop from the input source file is annotated by using the assigned loop identifier from the configuration and is written to a new file. 
Additionally, the sequence of loop transformations is added before the first loop. 
As an example, a loop nest such as 
\begin{minted}{c}
for (int i = 0; i < 128; i+=1)
  for (int j = 0; j < 128; j+=1)
    body(i,j);
\end{minted}
will be rewritten as the following.
\begin{minted}{c}
#pragma clang loop(loop1) parallelize_thread

#pragma clang loop id(loop1)
for (int i = 0; i < 128; i+=1)
  #pragma clang loop id(loop2)
  for (int j = 0; j < 128; j+=1)
    body(i,j);
\end{minted}

The path to the original file is replaced with the path to the rewritten file in the compiler command line. 
\textinline{-polly-output-loopnest} or debug symbols are not necessary this time, but \textinline{-fopenmp} is added in case any loop is parallelized.
Moreover, \textinline{-Werror=pass-failed} is added such that the compiler will abort with an error code when a transformation cannot be applied, instead of just warning about it.

If either the compilation or the execution fails, for example because of exceeding a timeout, the configuration is marked as invalid.
Doing so will avoid further exploration of ineffective transformations.
\Cref{fig:mctree_dot} shows such configurations in red.

% How time is taken
The time is measured as the wall-clock time of the entire executable. 
This includes overhead such as loading the executable and parts of the program that are not modified.
Hence, for our experiments we chose problem sizes such that the kernel execution time dominates the total wall-clock time.

The tool supports multiple loop nests in the application. 
A global configuration is the list of transformations for each loop nest. 
A new global configuration is derived from an existing configuration by applying one transformation to one of the loop nests.
However, in our experiments we only tune a single loop nest.

%%%%%%%%%%%%%%%%%%%%%%%%%%%%%%%%%%%%%%%%%%%%%%%%%%%%%%%%%%%%%%%%%%%%%%%%%%%%%%%%
\section{Evaluation}

We applied our tool to three  kernels selected from the PolyBench~\cite{polybench421} benchmark suites in the \cinline{EXTRALARGE_DATASET} size configuration using double precision.
The kernels are gemm, syr2k, and covariance.
Because loop distribution is not one of the supported transformations, we manually split loops to form larger perfectly nested loops that can be tiled and permuted.

We selected 4, 16, 64, 256, and 1024 (i.e., powers of 4) as the possible tile sizes, as a trade-off between approaching the best tile size vector and size of the search space.
For a perfect loop nest with three loops, this results in $5^3 + 2*5^2 + 3*5 = 190$ possibilities for tiling, $5!-1 = 5$ loop permutations, and $3$ configurations that parallelize one of the loops. 

One advantage  of using Polly is that it has no problem tiling and interchanging the nonrectangular loop nests in syr2k and covariance, which otherwise would be difficult to implement.
However, Polly does not consider the associativity or commutativity of the addition, thus limiting the permutations that are considered semantically legal, but also avoiding floating-point rounding differences.

The autotuning experiments were executed on a 2-socket Intel Xeon Platinum 8180M system with 376 GiB of main memory.
The cache sizes are 32 KiB for the L1d cache, 1 MiB for the L2, and 38.5 MiB for the L3 cache.
The autotuner was allowed to run for 6 hours, after which the job scheduling system would abort the job.

With 2x hyperthreading, a total of 112 threads are available. 
For each benchmark, we also separately autotuned without the parallelization transformation.
The choice of a platform with that many threads was intentional to show the effect on the search process.

%%%%%%%%%%%%%%%%%%%%%%%%%%%%%%%%%%%%%%%%%%%%%%%%%%%%%%%%%%%%%%%%%%%%%%%%%%%%%%%%
\section{Experimental Results}

The following graphs show the development of the autotuning process starting with the blue baseline experiment 0 without any loop transformations. 
Any successfully executed experiment has its execution time marked with a cross.
If it is faster than the previously fastest experiment, it is marked with a red cross and lowers the red \dquote{new best} bar.
That is, at the end of the graphs, the red line shows the overall best execution time compared with the starting point in blue.

\subsection{gemm}

PolyBench's gemm is a naive implementation of a generalized matrix-matrix multiplication kernel.
The extra-large configuration uses input matrices of sizes 2000x2600 and 2600x2300.

\begin{figure*}
\includegraphics[width=\textwidth]{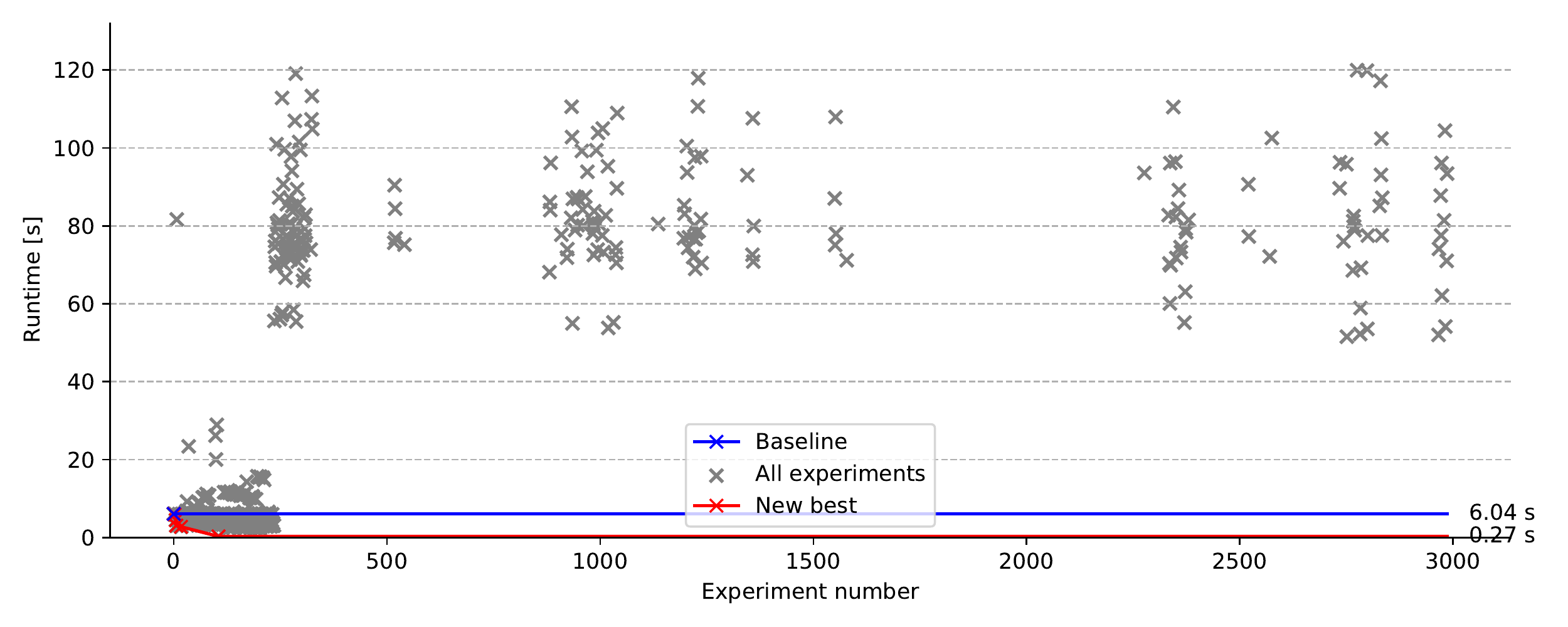}
\caption{Autotuning of gemm using tile/interchange/parallelize\_thread.}\label{fig:gemm_parallel}
\end{figure*}

\Cref{fig:gemm_parallel} visualizes the autotuning process where the best runtime is reached at experiment 105, which is the configuration
\begin{minted}{c}
#pragma clang loop(i) parallelize_thread
\end{minted}
i.e., the parallelization of the outermost loop.
We would have expected the best transformation to be parallelization after tiling/interchange that optimizes for cache accesses as our initial motivation does.
We can track this behavior to our search strategy: once the parallelization of the outermost loop is found, because of the high number of hardware threads no other sibling configuration comes close to its performance. 
By search space construction, a parallelized loop is not available for further transformations in derived configurations.
In the hierarchical configuration space, applying a different transformation, such as tiling, is a sibling configuration onto which parallelization can be applied in a second step.

Therefore, when parallelizing the outermost loop, its derived configurations only transforms the inner loops, which has a much smaller impact.
Since our search expands the leading configuration, the autotuner is stuck in a local minimum.

The same experiment without parallelization is shown in \cref{fig:gemm_nonparallel}. 
Its fastest configuration, shown below, resembles more our result from \cref{lst:gemmcore}.
\begin{minted}[fontsize=\footnotesize]{c}
#pragma clang loop(j,k,i) tile sizes(1024,64,16)
#pragma clang loop(i,j,k) interchange permutation(j,k,i)
\end{minted}
Its progress is also more interesting in that multiple new best configurations with different tile sizes are found before it settles on experiment number 381.
The difference of the baseline is due to noise in the measurement, which the baseline is most sensitive to because it initiates the warm-up. 
However, the found configurations and their runtime are consistent to the tens of milliseconds between multiple autotuning runs.
% FIXME: Too much noise!

\begin{figure*}
\includegraphics[width=\textwidth]{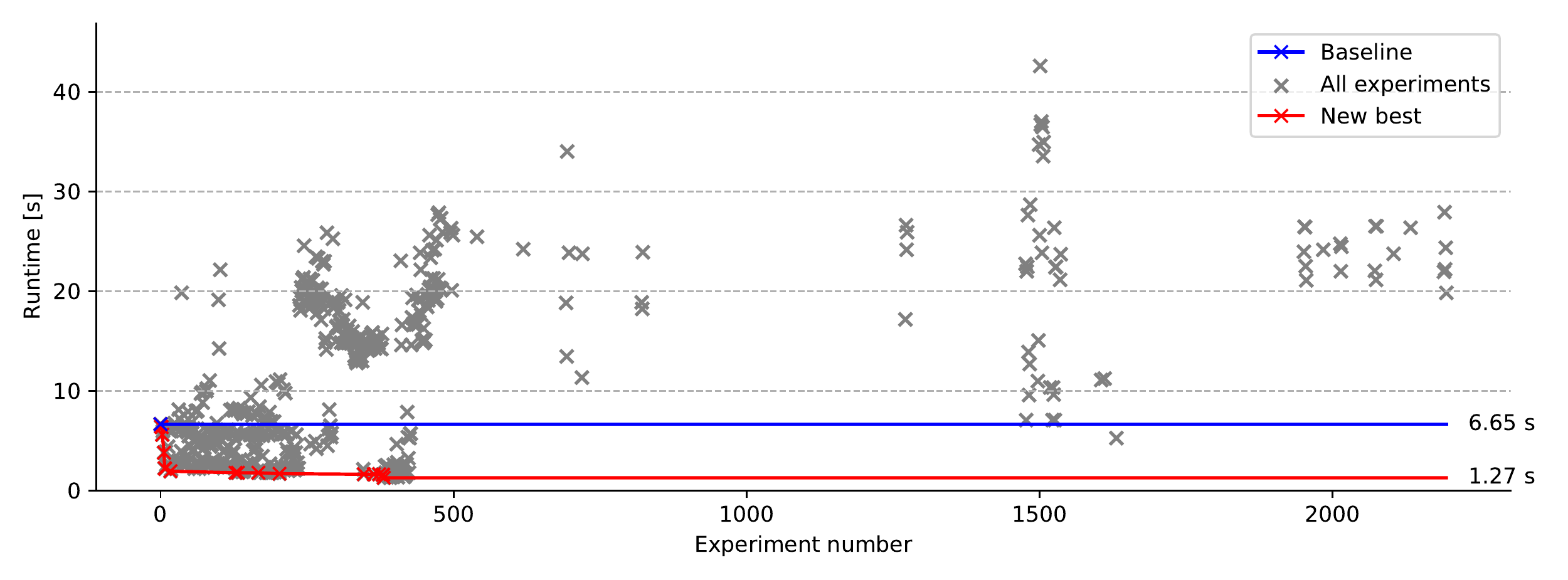}
\caption{Autotuning of gemm using tile/interchange.}\label{fig:gemm_nonparallel}
\end{figure*}
% TODO: Why is this faster than our analytical result?

Noticeably, the worst configurations with parallelization are three times slower than the worst configurations without parallelization. 
These results are due to the overhead introduced by parallelizing the innermost loop.

\subsection{syr2k}

syr2k is a symmetric rank-2k update BLAS routine. 
It is structurally similar to two matrix multiplications in a single loop nest but is nonrectangular.
Its extra-large configuration has input matrices of size 2600x3000.

\begin{figure*}
\includegraphics[width=\textwidth]{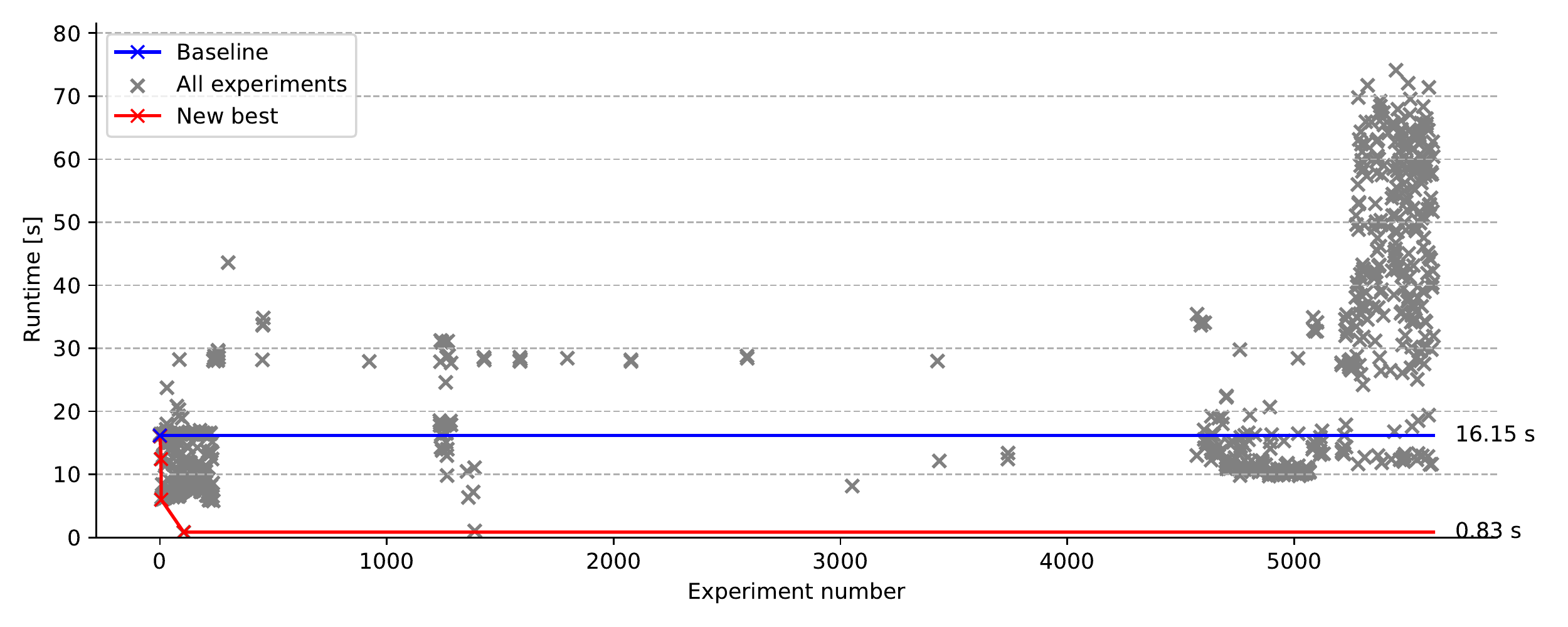}
\caption{Autotuning of syr2k using tile/interchange/parallelize\_thread.}\label{fig:syr2k_parallel}
\end{figure*}

Like gemm, the autotuning of syr2k (shown in \cref{fig:syr2k_parallel}) reaches the configuration that parallelizes the outermost in experiment 105 and then is trapped in a local minimum.

%\begin{figure*}
%\includegraphics[width=\textwidth]{figs/syr2k_nonparallel-36yx77yi}
%\caption{Autotuning of syr2k using tile/interchange}
%\end{figure*}

\begin{figure*}
\includegraphics[width=\textwidth]{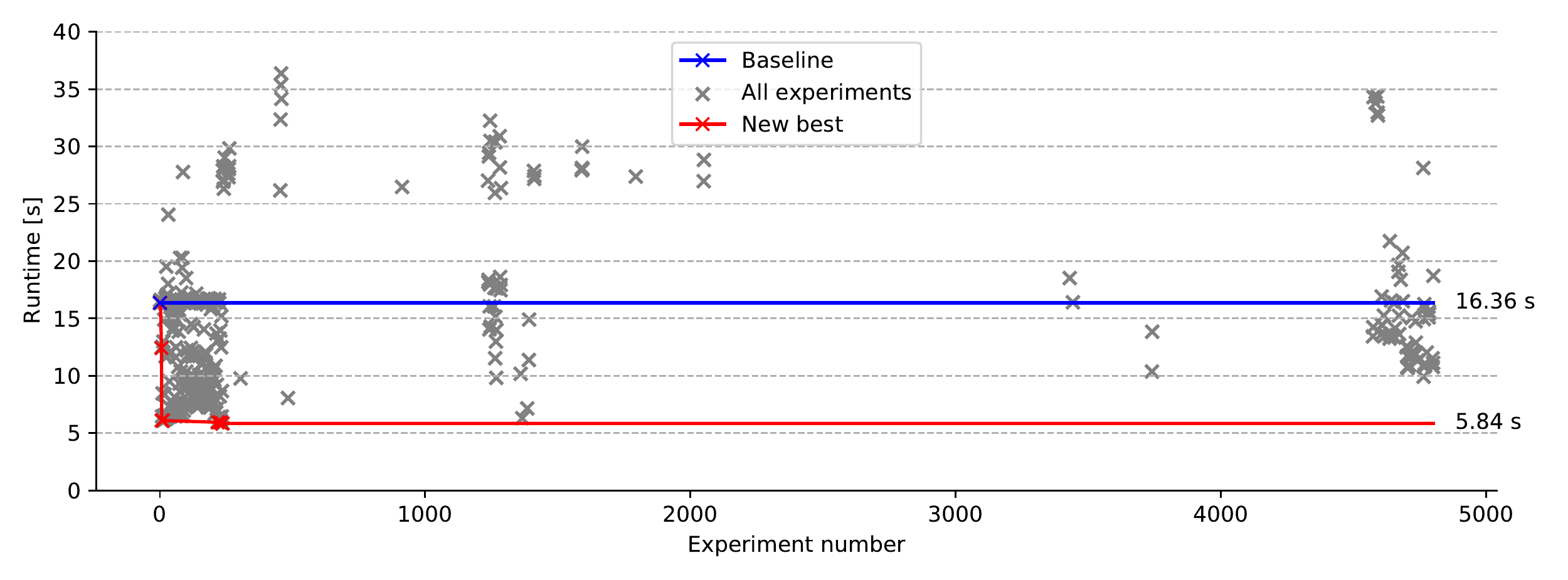}
\caption{Autotuning of syr2k using tile/interchange.}\label{fig:syr2k_nonparallel}
\end{figure*}

Without parallelization shown in \cref{fig:syr2k_nonparallel}, the best result is reached in experiment 105, which is
\begin{minted}{c}
#pragma clang loop(i,j,k) tile sizes(1024,16,64)
\end{minted}
The large number of unsuccessful configurations is due to the compiler's dependency check failing and thus not being able to ensure the semantic legality.

\subsection{covariance}

This benchmark computes the covariances between all columns of a matrix.
It consists of multiple sequentially executed loop nests of which the deepest also has a depth of 3 and is nonrectangular.
The input matrix size has the dimensions 3000x2600.

\begin{figure*}
\includegraphics[width=\textwidth]{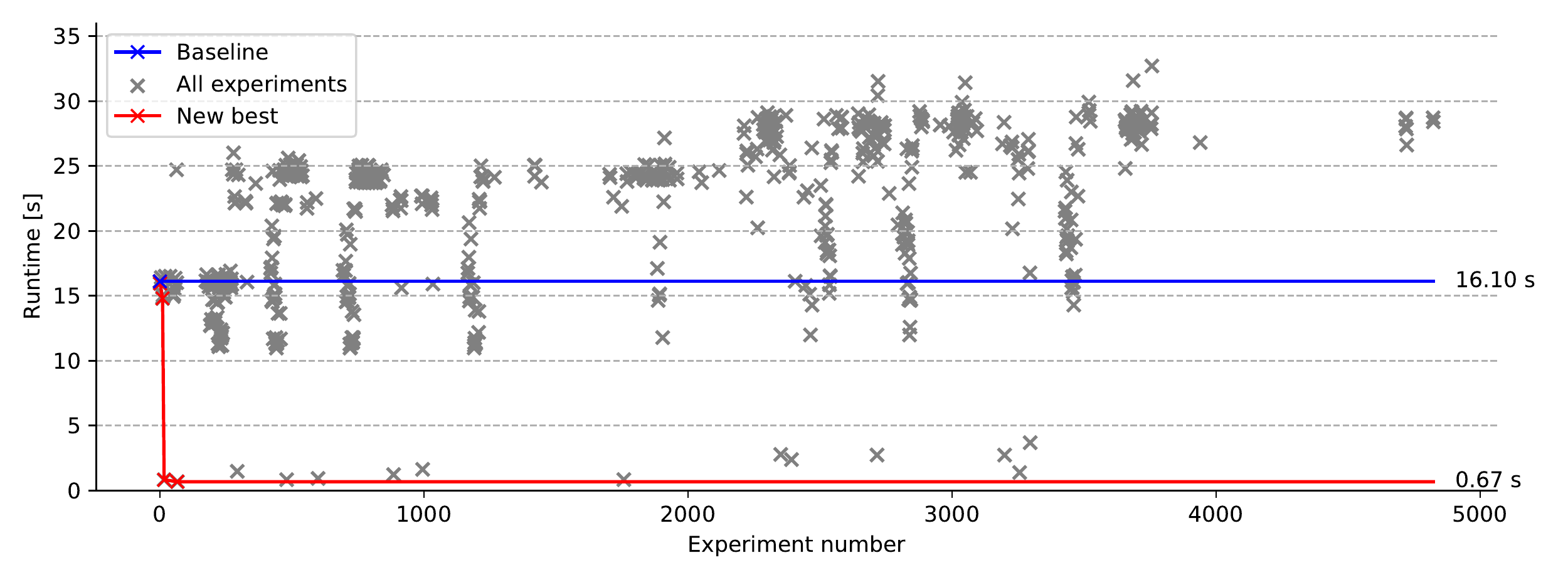}
\caption{Autotuning of covariance using tile/interchange/parallelize\_thread.}\label{fig:covariance_parallel}
\end{figure*}

The configuration that parallelizes the outermost loop is again found early in experiment 66 (\cref{fig:covariance_parallel}) and remains the best configuration until the end of the autotuning.

%\begin{figure*}
%\includegraphics[width=\textwidth]{figs/covariance_parallel-p726egyl.pdf}
%\caption{Autotuning of covariance using tile/interchange/parallel.}\label{fig:covariance_parallel}
%\end{figure*}

Without parallelization, multiple configurations using tiling are discovered until the best tile sizes
\begin{minted}{c}
#pragma clang loop(i,j,k) tile sizes(64,256,4)
\end{minted}
are found in experiment 225.
No faster configuration with more transformations  is found.

We note that this configuration is just two times slower than the massively parallel variant. 
This again shows that we missed profitable outer-level tiling opportunities before parallelizing the outermost loop.

\begin{figure*}
\includegraphics[width=\textwidth]{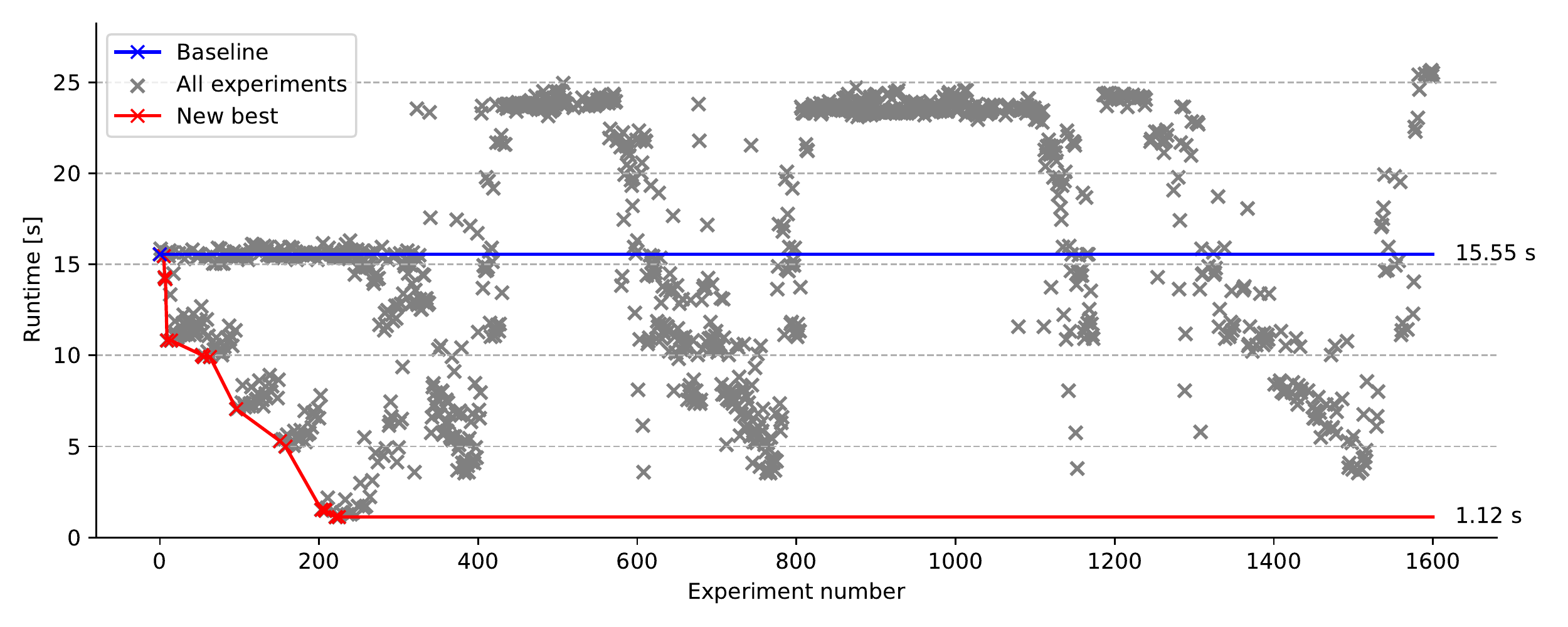}
\caption{Autotuning of covariance using tile/interchange.}\label{fig:covariance_nonparallel}
\end{figure*}

After finding a new best tiling, there is a noticeable cluster following with similar performance.
This is due to the configurations sharing the same or similar tile sizes for the inner loop, which seem to be most significant for the performance for this benchmark.

\section{Related Work}\label{sct:relatedwork}

Frameworks have been created to support autotuning independent of the domain, such as cTuning-cc/Milepost GCC~\cite{fursin11-milepost}, OpenTuner~\cite{ansel14-opentuner}, ATF~\cite{rasch17-atf}, and ytopt~\cite{ytopt}.
However, these tools expect the search space expressed in a vector space, that is, a fixed number of knobs. 
A search tree does not directly map to this space. 
%Only ATF also supports interdependent parameters.

Some of the approaches are based on a user-defined sequence of transformations where only the transformation parameters, such as tile sizes, are autotuned. 
CHiLL~\cite{chill}, atJIT~\cite{devmtg18-atjit}, and~\cite{sreenivasan18-surf} use this approach.
It still requires user interaction and therefore is only semi-automatic.

To allow more flexibility, one can add enable/disable transformation knobs to the search space.
Unfortunately, doing so either makes the knobs nonfunctional and thus enlarges the search space or, if the knobs are repurposed for other transformations, creates discontinuities in the performance evaluation. 
This breaks performance regression modeling approaches~\cite{pedro19-regression}. While other machine learning methods might handle this better, it increases the number of evaluations necessary to  discover that there even is a discontinuity.
We experimented with this approach ourselves using ytopt and Plopper~\cite{ytopt-plopper}.

A more extreme example of this approach is HalideTuner~\cite{zingales15-halidetuner}, which encodes a transformation sequence into a vector for OpenTuner.
Before evaluating a configuration, a verifier checks whether the vector encodes a valid loop transformation sequence.
ATF has a similar filtering concept called interdependent parameters.

The successor of HalideTuner, \cite{adams19-halidetree} uses tree search to avoid the limitation of a vector search space, but uses Beam search to explore the space. 
The publication reports much more efficient configuration generation than HalideTuner, and a factor two improvement over the more restricted default auto-scheduler that is integrated into Halide.
ProTuner~\cite{hajali20-protuner} further improves Halide schedule autotuning by replacing Beam search with MCTS.
MCTS find schedules which are up to 3.6 times faster than the best found by Beam search, in part because of Beam search's non-global behavior.

Instead of explicit transformations, LeTSeE~\cite{pouchet08-letsee} autotunes a scatter function, commonly used in polyhedral optimization, which is also used by Polly.
A scatter function assigns statement executions to points in time.
Assigning two executions to the same time means execution in parallel.
This representation has no artificial discontinuities, and the legality of a configuration is ensured by adding boundaries to the search space.
However, not all transformations (such as vectorization, unrolling, outer parallelism) can be represented in this model.

Autotuners that specialize on a specific class of algorithms also exist. 
These typically have a built-in knowledge of which transformations to apply and how to tune the parameters.
For instance, PATUS~\cite{christen11-patus} and an extension of LIFT~\cite{hagedorn18-lift} optimize stencils generated from their domain-specific languages.
LIFT uses ATF as its autotuning framework.
The main parameter it tunes for is the tile size to fit the memory hierarchy, but they also allow for domain-specific optimization techniques. 

Of these related works, only~\cite{sreenivasan18-surf} and~\cite{ytopt-plopper} also use compiler pragmas to apply transformations.
Most other systems use source-to-source transformations or code generation to apply transformations themselves.
In contrast to our solution which requires a compiler extension, these are compatible with any compiler.
Because of this lack of portability, we are working on standardizing loop transformation as part of OpenMP~\cite{iwomp18-pragmas}.
Unrolling and tiling will be part of OpenMP 5.1, with more intended to be added in OpenMP 6.0.
OpenMP as a search domain is used by, for example, \cite{sreenivasan18-surf}.

%%%%%%%%%%%%%%%%%%%%%%%%%%%%%%%%%%%%%%%%%%%%%%%%%%%%%%%%%%%%%%%%%%%%%%%%%%%%%%%%
\section{Future Work}

% Search space exploration improvements
While our autotuner is meant only  to demonstrate the search space generation for composable loop transformations, considerable research has been done on search space exploration for autotuning~\cite{balaprakash18-surf,ashouri18-survey}, which we could apply to our autotuner.

A case can be made for using different algorithms for the exploration of composing loop transformations and their parameters.
The parameters often have numeric relations for which specialized frameworks such as ytopt~\cite{ytopt} already exist.
However, there is less autotuning support for search spaces with tree or DAG structure.
Here, our original idea was to implement a Monte Carlo tree search (MCTS), and our intention was to integrate it into the ytopt framework to exploit both kinds of search spaces.
This could take the form of nodes in the MCTS representing a vector search space of the parameters used by the transformations in that node.
The machine learning algorithm's job then is to balance exploring a parameter search space and expanding a new MCTS node.
Which nodes to explore could also be influenced by the improvements adding a specific transformation had on other configurations.

Equal configurations reachable through different paths through the search space should be detected and merged into the same node,  such that search space becomes a directed acyclic graph.
Adding transformations before a transformation already in the configuration would likely allow escaping the local minimum of only paralleling the innermost loop but would make the search space even less treelike.

Instead of measuring the execution time of the entire executable, the source code rewriter could insert profiling instructions to measure the execution time of each loop nest individually.
We also intend to implement more transformations, such as loop distribution, unrolling, vectorization, and maybe even accelerator offloading.

%%%%%%%%%%%%%%%%%%%%%%%%%%%%%%%%%%%%%%%%%%%%%%%%%%%%%%%%%%%%%%%%%%%%%%%%%%%%%%%%
\section{Conclusion}

We demonstrated the viability of the autotuning search space for loop transformations that has the straightforward representation as either a tree or a directed acyclic graph.
The loop transformations are applied by using our implementation of user-directed loop transformation in Clang and Polly. 
A simple autotuner implementation is able to derive new loop transformation configurations, insert them into the source code as pragmas, compile, and execute the modified program to determine its performance.

Our exploitation-only strategy considering only the fastest configuration so far proved to be the weak point in our experiments on three selected PolyBench benchmarks: it tends to stay in the local minimum of just parallelizing the outermost loop and then explore only less-significant transformations of its nested loops.
Without parallelization, however, it found highly effective tiling and loop interchange configurations.
Unfortunately, it did not find multilevel tiling strategies for each cache in the memory hierarchy.
The tree expansion would need to be more directed by using known relations rather than broad evaluation of all tiling sizes.

We intend to continue to improve the search space generation by integration into the ytopt framework using a hybrid of Monte Carlo tree search exploration strategies and machine learning on vector spaces.

Using pragmas inserted into the source code by the autotuner has several advantages.
First, we can use compiler infrastructure to determine whether a transformation is guaranteed to preserve the program's semantics (in contrast to, e.g., comparing the output of a program to a reference result, which does not guarantee that all runs give correct results).
Second, one can take the modified source of the best configuration for use in production.
Third, the modified source code remains understandable and can be modified further, either for manual tweaking of the result or for further  application development.
While we currently rely on our own implementation of loop transformation pragmas in Clang, the addition of similar directives into the OpenMP standard will allow this kind of autotuning with all OpenMP-supporting compilers.

%%%%%%%%%%%%%%%%%%%%%%%%%%%%%%%%%%%%%%%%%%%%%%%%%%%%%%%%%%%%%%%%%%%%%%%%%%%%%%%%
\section*{Acknowledgment}

This research was supported by the Exascale Computing Project (17-SC-20-SC), a collaborative effort of the U.S. Department of Energy Office of Science and the National Nuclear Security Administration, in particular its subproject PROTEAS-TUNE.

This research used resources of the Argonne Leadership Computing Facility, which is a DOE Office of Science User Facility supported under Contract DE-AC02-06CH11357.

This material was based in part upon funding from the U.S. Department of Energy, Office of Science, under contract DE-AC02-06CH11357.

\IEEEtriggeratref{14}
\bibliographystyle{IEEEtran}
\bibliography{bibliography}

\end{document}